\documentclass{kluwer}
\usepackage[dvips]{graphicx}
\newdisplay{guess}{Conjecture}

\begin{document}                                                                                   
\begin{article}
\begin{opening}         
\title{On the multiplicity of the O-star Cyg\,OB2\,\#8A and its contribution to the $\gamma$-ray source 3EG J2033+4118\thanks{Based partly on data obtained at the Observatoire de Haute-Provence, France.}} 
\author{Michael \surname{De Becker}, Gregor \surname{Rauw} and Jean-Pierre \surname{Swings}}  
\runningauthor{De Becker et al.}
\runningtitle{On the multiplicity of the O-star Cyg\,OB2\,\#8A}
\institute{Institut d'Astrophysique et de G\'eophysique, Universit\'e de Li\`ege, 17, All\'ee du 6 Ao\^ut, B5c, B-4000 Sart Tilman, Belgium}
\date{July 27, 2004}

\begin{abstract}
We present the results of an intensive spectroscopic campaign in the optical waveband revealing that Cyg\,OB2\,\#8A is an O6 + O5.5 binary system with a period of about 21.9 d. Cyg\,OB2\,\#8A is a bright X-ray source, as well as a non-thermal radio emitter. We discuss the binarity of this star in the framework of a campaign devoted to the study of non-thermal emitters, from the radio waveband to $\gamma$-rays. In this context, we attribute the non-thermal radio emission from this star to a population of relativistic electrons, accelerated by the shock of the wind-wind collision. These relativistic electrons could also be responsible for a putative $\gamma$-ray emission through inverse Compton scattering of photospheric UV photons, thus contributing to the yet unidentified EGRET source 3EG J2033+4118.
\end{abstract}
\keywords{binaries: spectroscopic -- stars: individual: Cyg\,OB2\,\#8A -- stars: early-type -- radiation mechanisms: non-thermal}
\end{opening}           

\section{Introduction}
Cyg\,OB2\,\#8A (BD\,+40$^\circ$\,4227, m$_\mathrm{V}$=9.06) is one of the optically brightest O-stars in Cyg\,OB2. This star is classified as O5.5I(f) \cite{MT}, and is known to be a bright non-thermal and strongly variable radio emitter \cite{BAC}. Up to now, Cyg\,OB2\,\#8A had never been shown to be a binary system.

However, in the context of the multiwavelength study of non-thermal radio emitters, the binarity is expected to play a crucial role. Indeed, as described in this communication, the most plausible scenario that could explain the non-thermal emission from hot stars relies on the existence of shocks able to accelerate electrons up to relativistic energies. Whether these shocks are intrinsic to the wind or due to a wind-wind collision in a binary system remains an open question. For this reason, we initiated an optical monitoring of non-thermal emitting massive stars. The result we report here is the discovery that Cyg\,OB2\,\#8A is a binary system. In this context, we discuss the expected emission of X and $\gamma$-rays from its interacting wind zone, and its contribution to the unidentified EGRET source 3EG J2033+4118 \cite{ben}.

This communication is organized as follows. Section \ref{binary} discusses the binarity of Cyg\,OB2\,\#8A and presents its first orbital solution. Section \ref{arc} deals with archive X-ray data in the context of a binary scenario. A discussion of non-thermal processes in the framework of massive stars is given is Section \ref{ntsect}, and the conclusions appear in Section \ref{conc}.

\section{Cyg\,OB2\,\#8A as a binary system \label{binary}}
\subsection{Spectroscopic analysis}
We obtained a series of 35 spectra in the blue range (between about 4455 and 4900 \AA\,) at the Observatoire de Haute-Provence (OHP, France) during four observing runs in September 2000, September 2001, September 2002 and October 2003. For details on the observations and the reduction procedure, see \inlinecite{let}.

\begin{figure}
\centerline{\includegraphics[width=18pc]{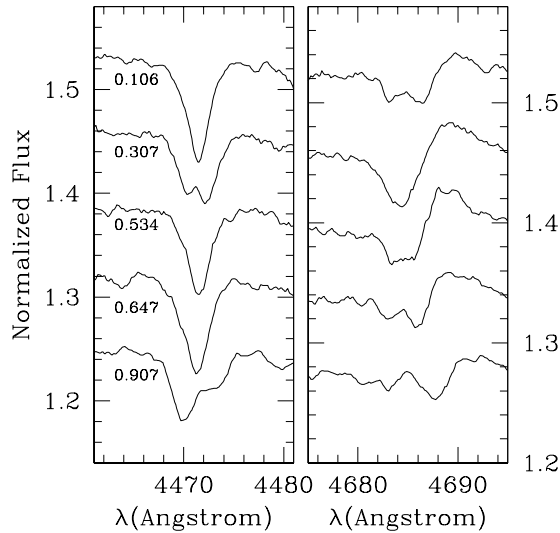}}
\caption{Profiles of the HeI $\lambda$ 4471 (left) and HeII $\lambda$ 4686 (right) lines displayed at 5 different phases specified in the left panel. (Preliminary results from De Becker \& Rauw, 2004).}
\label{prof}
\end{figure}

Our spectra clearly indicate that Cyg\,OB2\,\#8A is a binary system consisting of an O6 primary and an O5.5 secondary revolving around each other in {\bf 21.908 days} \cite{let}. The intensities of the lines in the spectra of the two components roughly yield a visual brightness ratio of 2 between the primary and the secondary (see Fig.\,\ref{prof}). Since the spectral types of both components are similar, this suggests that their luminosity classes might be different.

Fig.\,\ref{prof} shows some profiles of the HeI $\lambda$ 4471 and HeII $\lambda$ 4686 lines. We clearly see that the HeII line displays a strong variability. Some phases show P-Cygni profiles ($\phi$\,=\,0.307), but at other phases the profiles are much more complicated (e.g. $\phi$\,=\,0.907). The observed variations are phase-locked and most probably reflect an interaction between the stellar winds of the two components \cite{pro}.

\subsection{Orbital solution \label{os}}
By simultaneously fitting two Gaussians to the HeI $\lambda$ 4471 line profiles, we determined the radial velocities of the two stars. We obtained the first orbital solution for this system using the method described by \inlinecite{sana}. We assigned various weights to our data to take into account the different errors affecting our RV measurements. Due to the intensity ratio of the lines of the two components ($\sim$ 2), there was no ambiguity on the identification of the lines respectively of the primary and the secondary. Table\,\ref{par} yields the main parameters of the system \cite{let}. However, we emphasize that the errors on the eccentricity and the period probably underestimate the actual error as a result of the rather heterogeneous phase coverage of our time series.\\

\begin{figure}
\centerline{\includegraphics[width=18pc]{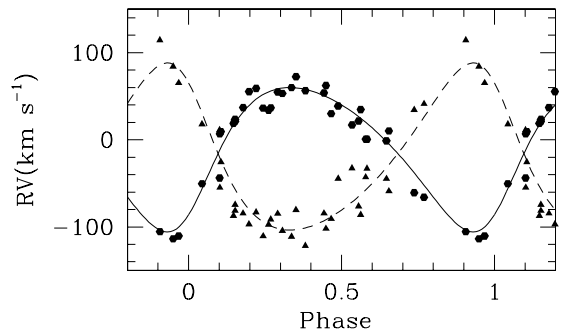}}
\caption{Radial velocity (RV) curve of Cyg\,OB2\,\#8A for an orbital period of 21.908 d. The hexagons (resp. triangles) stand for the primary (resp. secondary) RVs. The solid and dashed lines yield our best fit orbital solution respectively for the primary and the secondary (from De Becker et al.,\,2004a).}
\label{orb}
\end{figure}

\begin{table}
\caption{Orbital solution for Cyg\,OB2\,\#8A. T$_{\circ}$ refers to the time of periastron passage. $\gamma$, $K$, and $a\,\sin\,i$ denote respectively the systemic velocity, the amplitude of the radial velocity curve, and the projected separation between the centre of the star and the centre of mass of the binary system.}
\label{par}
\begin{tabular}{l c c }
\hline
  & Primary & Secondary \\
\hline
P (days)  & \multicolumn{2}{c}{21.908 (fixed)} \\
$e$   & \multicolumn{2}{c}{0.24 $\pm$ 0.04} \\
T$_\circ$ (HJD--2\,450\,000) & \multicolumn{2}{c}{1807.139 $\pm$ 0.894} \\
$\gamma$ (km\,s$^{-1}$)  & --8.1 $\pm$ 3.3 & --25.0 $\pm$ 3.6 \\
$K$ (km\,s$^{-1}$) & 82.8 $\pm$ 3.5 & 95.8 $\pm$ 4.0 \\
$a\,\sin\,i$ (R$_\odot$) & 34.8 $\pm$ 1.5 & 40.3 $\pm$ 1.7 \\
$m\,\sin^3i$ (M$_\odot$)  & 6.4 $\pm$ 0.6 & 5.5 $\pm$ 0.5 \\
\hline
\end{tabular}
\end{table}

\section{Archive X-ray data \label{arc}}
\begin{figure}
\centerline{\includegraphics[width=18pc]{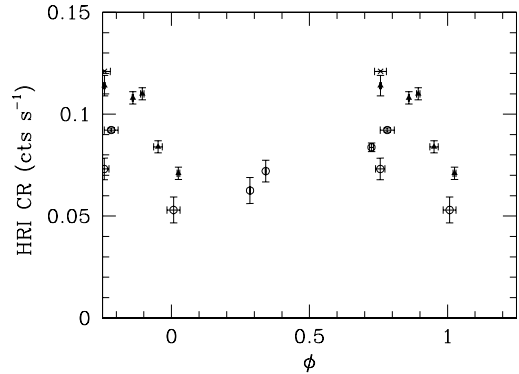}}
\caption{Equivalent {\it ROSAT}/HRI count rates folded with the ephemeris taken from Table\,\ref{par}. Filled triangles, open circles and the cross stand respectively for {\it ROSAT}/PSPC, {\it ROSAT}/HRI and {\it ASCA}/SIS data.}
\label{arch}
\end{figure}

Cyg\,OB2\,\#8A archive data obtained with the {\it ROSAT} X-ray satellite were analyzed. For the pointings obtained with the PSPC (Position Sensitive Proportional Counter) instrument, we extracted spectra which were fitted with optically thin thermal plasma models \cite{mekal}. After convolution with the HRI (High Resolution Imager) response matrix we estimated the equivalent HRI count rates. These results are quoted as filled triangles in Fig.\,\ref{arch} that yields the archive X-ray count rates folded with the orbital parameters given in Table\,\ref{par}. The 6 open circles correspond to observations performed with the HRI.

This star was also observed with the {\it ASCA} satellite. In that case, a complete spectral analysis of the SIS (Solid-state Imaging Spectrometer) data was performed \cite{master}. The X-ray spectrum was successfully fitted with an absorbed two-temperature optically thin plasma model \cite{mekal}, with temperatures of about 8\,$\times$\,10$^{6}$ and 2\,$\times$\,10$^{7}$ K respectively for the two components. The high temperature of the second component is suggestive of a plasma heated by a wind-wind interation within a binary system \cite{SBP}. The cross in Fig.\,\ref{arch} quotes the equivalent {\it ROSAT}/HRI count rate of this {\it ASCA} observation.

Archive {\it Einstein} data were also retrieved, but a first examination revealed important discrepancies, probably partly due to different locations of the source in the IPC (Imaging Proportional Counter) detector plane. For this reason, we did not include these data in our archive X-ray data investigation.  

Although the X-ray light curve is too scarce to draw firm conclusions, Fig.\,\ref{arch} clearly reveals that there are important variations in the X-ray flux of this system. We see that the PSPC and SIS data point to a high emission level at phases just before periastron. Then, the count rate apparently decreases. This trend is compatible with an interacting wind scenario, leading to a higher X-ray flux when the stars are closer. The fact that the maximum does not coincide with the minimum separation could possibly be explained by the fact that, at these phases, the winds have not reached their terminal velocities before they collide. Variable circumstellar absorption is also expected to play a role in the shape of this curve. However, these trends are less clear if we consider the HRI data. This X-ray lightcurve will be confronted to the results of several {\it XMM-Newton} pointings scheduled for the end of 2004.

\section{Non-thermal emission from massive stars \label{ntsect}}
\subsection{General scenario}
The non-thermal radio emission from massive stars raises some issues in the context of massive star physics. First, it reveals the existence of a mechanism that accelerates electrons up to relativistic energies. This acceleration could be achieved through the first order Fermi mechanism, believed to be responsible for, e.g., the production of cosmic particles (\opencite{Bella}; \citeyear{Bellb}). Second, the identification of the non-thermal radio emission from several early-type stars as synchrotron emission by \inlinecite{wh} implies that a magnetic field must be present in their winds.

\begin{figure}
\centerline{\includegraphics[width=27pc]{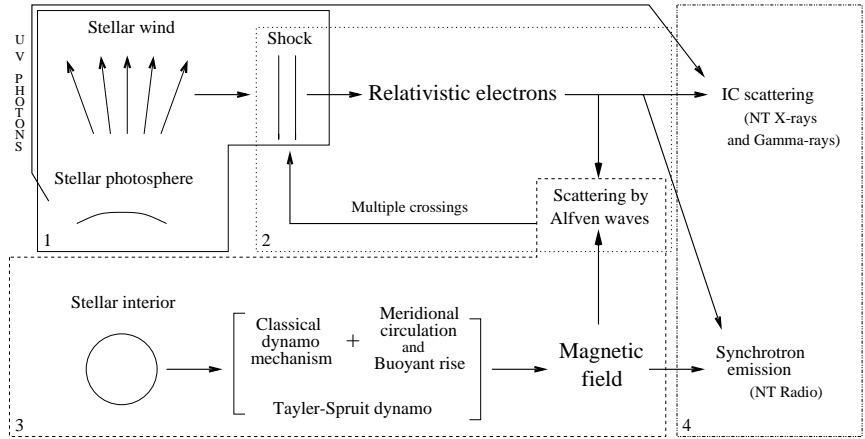}}
\caption{General scenario proposed for the production of the non-thermal emission of massive stars in the radio, X-ray and $\gamma$-ray wavebands.}
\label{nt}
\end{figure}

Figure \ref{nt} summarizes the main aspects of the general scenario currently proposed to produce non-thermal emission in the context of massive stars. Box 1 schematically illustrates the physics of stellar winds, from the production of a stellar wind through the radiation pressure \cite{cak} to the formation of hydrodynamic shocks through intrinsic instabilities \cite{feld} or wind-wind collision in a binary systems \cite{SBP}. Box 2 illustrates the Fermi acceleration mechanism occuring within shocks, with the support of Alfv\'en waves generated by magnetohydrodynamic phenomena. The magnetic field of a massive star (box 3) is expected either to be produced through a classical dynamo mechanism inside the convective core, and to travel up to the surface with the support of meridional circulation \cite{CM} and/or buoyancy \cite{MC}, or through a Tayler-Spruit type dynamo driven by the differential rotation in the non-turbulent radiative zone \cite{Spr}. Finally, box 4 illustrates the non-thermal emission of radiation in the radio waveband, as well as in the X-ray and $\gamma$-ray domains (the latter two through inverse Compton (IC) scattering). The intersections between boxes illustrate the noticeable interconnections between these different fields of astrophysics.  

\subsection{Cyg\,OB2\,\#8A in the context of this scenario}
As Cyg\,OB2\,\#8A is now a confirmed binary, we can expect that the interaction zone between the two winds is responsible for the acceleration of the relativistic electrons that produce the radio synchrotron emission \cite{EU}. Therefore, the same region could be at the origin of a high energy counterpart of this non-thermal radio emission. Indeed, IC scattering of UV photospheric photons could produce X-rays and $\gamma$-rays \cite{CW}. For a general discussion of non-thermal emission from early-type binaries, see \citeauthor{rauw} (\citeyear{rauw}, this conference).

In the X-rays, the strong thermal emission due to the colliding winds prevents us from clearly detecting a (putative) non-thermal emission component, like in the case of the non-thermal radio emitter HD\,168112 studied by \inlinecite{DeB2}. However, in the $\gamma$-ray domain, no thermal emission mechanism is expected and the detection of an object like Cyg\,OB2\,\#8A in this energy domain would provide unambigous evidence for a high-energy counterpart to the non-thermal radio emission. For this reason, we obtained observing time with the European {\it INTEGRAL} satellite. With these observations, we intend to detect the counterpart of the EGRET source 3EG J2033+4118 \cite{ben} and to evaluate the contribution of Cyg\,OB2\,\#8A to this yet unidentified $\gamma$-ray source. Following the study of \inlinecite{hart}, Cyg\,OB2\,\#8A lies within the 99\,\% probability contour of this EGRET source.

We estimated the expected IC luminosity (L$_{IC}$) following the same approach as \inlinecite{ben}. For Cyg\,OB2\,\#8A, we adopted a radial dependence of the magnetic field \cite{EU}. The collision zone is indeed closer to the stars than in the case considered by \inlinecite{ben}. Adopting typical radii for O6I and O5.5III stars \cite{HP}, a surface magnetic field of 1 Gauss, a bolometric luminosity of 10$^{6.3}$\,L$_\odot$ and a synchrotron luminosity of 3.7\,$\times$\,10$^{28}$ erg\,s$^{-1}$ \cite{BAC}, we obtain an orbit averaged L$_{IC}$ of $\sim$\,2-3\,$\times$\,10$^{34}$\,erg\,s$^{-1}$ for a range of orbital inclinations of 30$^\circ$-45$^\circ$. A surface magnetic field of $\sim$\,100 G would give a L$_{IC}$ a factor 10$^{4}$ lower. For comparison, \inlinecite{ben} derive a value of about 8\,$\times$\,10$^{34}$\, erg\,s$^{-1}$ for Cyg\,OB2\,\#5. We also estimated the maximum energy that can be reached by the relativistic electrons under these conditions, using the relation given by \inlinecite{ben2}. A lower limit of the maximum lorentz factor should be about 10$^{4}$. This should lead to photons up to a few GeV. For a detailed discussion of the expected IC luminosity of Cyg\,OB2\,\#8A, see \inlinecite{pro}. 

\section{Prospects and conclusions \label{conc}}
We presented the results of an optical campaign revealing that the early-type star Cyg\,OB2\,\#8A is an O6 + O5.5 binary system likely seen under a relatively low inclination angle, with a period of about 21.9\,d and an eccentricity of 0.24. The binarity of this star is discussed in the context of the X-ray emission through archive data.

The binarity of this non-thermal radio emitter is of fundamental importance in the framework of our multiwavelength campaign devoted to the non-thermal emission from massive stars (\opencite{9sgr}; \opencite{DeB2}). The acceleration of relativistic electrons responsible for non-thermal emission (radio, X-rays, $\gamma$-rays) is believed to occur within the interaction zone of the colliding winds of the binary system \cite{EU}. The IC scattering of UV photons of Cyg\,OB2\,\#8A is expected to account for 8-10 \% of the $\gamma$-ray luminosity of the unidentified EGRET source 3EG J2033+4118. Forthcoming observations with the {\it INTEGRAL} satellite should allow to unambigously detect the lower energy $\gamma$-ray emission of the Cyg\,OB2 stars and then help to constrain the general scenario of non-thermal radiation from massive stars.

\acknowledgements
We are greatly indebted to the FNRS (Belgium) for multiple support including the rent of the OHP telescope in 2000 and 2002 through contract 1.5.051.00 "Cr\'edit aux chercheurs". The travels to OHP were supported by the Minist\`ere de l'Enseignement Sup\'erieur et de la Recherche de la Communaut\'e Fran\c{c}aise. This research is also supported in part by contract PAI P5/36 (Belgian Federal Science Policy Office) and through the PRODEX XMM-OM and INTEGRAL Projects.

\end{article}

\begin{thebibliography}{}
\bibitem[\protect\citeauthoryear{Bell}{1978a}]{Bella}
Bell, A.R.\ 1978a, MNRAS, 182, 147
\bibitem[\protect\citeauthoryear{Bell}{1978b}]{Bellb}
Bell, A.R.\ 1978b, MNRAS, 182, 443
\bibitem[\protect\citeauthoryear{Benaglia et al.}{2001}]{ben}
Benaglia, P., Romero, G.E., Stevens, I.R., \& Torres, D.F.\ 2001, A\&A, 366, 605
\bibitem[\protect\citeauthoryear{Benaglia \& Romero}{2003}]{ben2}
Benaglia, P., \& Romero, G.E.\ 2003, A\&A, 399, 1121
\bibitem[\protect\citeauthoryear{Bieging et al.}{1989}]{BAC}
Bieging, J.H., Abbott, D.C., \& Churchwell, E.\ 1989, ApJ, 340, 518
\bibitem[\protect\citeauthoryear{Castor et al.}{1975}]{cak}
Castor, J.I., Abbott, D.C., \& Klein, R.I.\ 1975, ApJ, 195, 157
\bibitem[\protect\citeauthoryear{Charbonneau \& MacGregor}{2001}]{CM}
Charbonneau, P., \& MacGregor, K.B.\ 2001, ApJ, 559, 1094
\bibitem[\protect\citeauthoryear{Chen \& White}{1994}]{CW}
Chen, W., \& White, R.L.\ 1994, Ap\&SS, 221, 259
\bibitem[\protect\citeauthoryear{De Becker}{2001}]{master}
De Becker, M.\ 2001, Master thesis, University of Li\`ege
\bibitem[\protect\citeauthoryear{De Becker \& Rauw}{2004}]{pro}
De Becker, M., \& Rauw, G.\ 2004, A\&A, in preparation
\bibitem[\protect\citeauthoryear{De Becker et al.}{2004a}]{let}
De Becker, M., Rauw, G., \& Manfroid, J.\ 2004a, A\&A, submitted
\bibitem[\protect\citeauthoryear{De Becker et al.}{2004b}]{DeB2}
De Becker, M., Rauw, G., Blomme, R., et al.\ 2004b, A\&A, 420, 1061
\bibitem[\protect\citeauthoryear{Eichler \& Usov}{1993}]{EU}
Eichler, D., \& Usov, V.\ 1993, ApJ, 402, 271
\bibitem[\protect\citeauthoryear{Feldmeier et al.}{1997}]{feld}
Feldmeier, A., Puls, J., \& Pauldrach, A.W.A.\ 1997, A\&A, 322, 878
\bibitem[\protect\citeauthoryear{Hartman et al.}{1999}]{hart}
Hartman, R.C., Bertsch, D.L., Bloom, S.D., et al.\ 1999, ApJS, 123, 79
\bibitem[\protect\citeauthoryear{Howarth \& Prinja}{1989}]{HP}
Howarth, I.D., \& Prinja, R.K.\ 1989, ApJS, 69, 527
\bibitem[\protect\citeauthoryear{MacGregor \& Cassinelli}{2003}]{MC}
MacGregor, K.B., \& Cassinelli, J.P.\ 2003, ApJ, 586, 480
\bibitem[\protect\citeauthoryear{Massey \& Thompson}{1991}]{MT}
Massey, P., \& Thompson, A.B.\ 1991, AJ, 101, 1408
\bibitem[\protect\citeauthoryear{Mewe et al.}{1987}]{mekal}
Mewe, R., Gronenschild, E.H.B.M., \& van den Oord, G.H.J.\ 1985, A\&AS, 62, 197
\bibitem[\protect\citeauthoryear{Rauw}{2004}]{rauw}
Rauw, G.\ 2004, this conference
\bibitem[\protect\citeauthoryear{Rauw et al.}{2002}]{9sgr}
Rauw, G., Blomme, R., Waldron, W.L., et al.\ 2002, A\&A, 394, 993
\bibitem[\protect\citeauthoryear{Sana et al.}{2003}]{sana}
Sana, H., Hensberge, H., Rauw, G., \& Gosset, E.\ 2003, A\&A, 405, 1063
\bibitem[\protect\citeauthoryear{Spruit}{2002}]{Spr}
Spruit, H.C.\ 2002, A\&A, 381, 923
\bibitem[\protect\citeauthoryear{Stevens et al.}{1992}]{SBP}
Stevens, I.R., Blondin, J.M., \& Pollock, A.M.T.\ 1992, ApJ, 386, 265
\bibitem[\protect\citeauthoryear{White}{1985}]{wh}
White, R.L.\ 1985, ApJ, 289, 698
\end{thebibliography}
\end{document}